\documentclass[%
 aip,
 amsmath,amssymb,
 reprint,%
]{revtex4-1}

\usepackage{graphicx}% Include figure files
\usepackage{dcolumn}% Align table columns on decimal point
\usepackage{bm}% bold math
\usepackage[utf8]{inputenc}
\usepackage[T1]{fontenc}
\usepackage{mathptmx}
\usepackage{etoolbox}
 \usepackage[normalem]{ulem} 
 \usepackage[dvipsnames]{xcolor}

\newcommand*{\sSG}{\sigma_{\scriptscriptstyle\rm SG}}

\newcommand*{\sGW}{\sigma_{\scriptscriptstyle\rm GW}}
\newcommand*{\sGO}{\sigma_{\scriptscriptstyle\rm GO}}
\newcommand*{\sSW}{\sigma_{\scriptscriptstyle\rm SW}}
\newcommand*{\sSO}{\sigma_{\scriptscriptstyle\rm SO}}
\newcommand*{\sWO}{\sigma_{\scriptscriptstyle\rm WO}}

\makeatletter
\def\@email#1#2{%
 \endgroup
 \patchcmd{\titleblock@produce}
  {\frontmatter@RRAPformat}
  {\frontmatter@RRAPformat{\produce@RRAP{*#1\href{mailto:#2}{#2}}}\frontmatter@RRAPformat}
  {}{}
}%
\makeatother
\begin{document}

\preprint{AIP/123-QED}

\title[Optimizing oil-water separation using fractal surfaces]{Optimizing oil-water separation using fractal surfaces}
% Force line breaks with \\

\author{Cristina Gavazzoni} 
\email{crisgava@gmail.com}
\affiliation{Instituto de F\'isica, Universidade Federal
do Rio Grande do Sul, Caixa Postal 15051, CEP 91501-970, 
Porto Alegre, Rio Grande do Sul, Brazil}

\author{Davi Lazzari} 
\email{lazzari26@hotmail.com}
\affiliation{Instituto de F\'isica, Universidade Federal
do Rio Grande do Sul, Caixa Postal 15051, CEP 91501-970, 
Porto Alegre, Rio Grande do Sul, Brazil}

\author{Iara Patrícia da Silva Ramos} 
\email{iarapaty24@gmail.com}
\affiliation{Instituto de F\'isica, Universidade Federal
do Rio Grande do Sul, Caixa Postal 15051, CEP 91501-970, 
Porto Alegre, Rio Grande do Sul, Brazil}

\author{Carolina Brito}
\email{carolina.brito@ufrgs.br}
\affiliation{Instituto de F\'isica, Universidade Federal
do Rio Grande do Sul, Caixa Postal 15051, CEP 91501-970, 
Porto Alegre, Rio Grande do Sul, Brazil}

\date{\today}% It is always \today, today,

\begin{abstract}
Oil has become a prevalent global pollutant, stimulating the research to improve the techniques to separate oil from water. Materials with special wetting properties—primarily those that repel water while attracting oil—have been proposed as suitable candidates for this task. However, one limitation in developing efficient substrates is the limited available volume for oil absorption.  In this study we investigate the efficacy of disordered fractal materials in addressing this challenge, leveraging their unique wetting properties. Using a combination of a continuous model and Monte Carlo simulations, we characterize the hydrophobicity and oleophilicity of substrates created through ballistic deposition (BD). Our results demonstrate that these materials exhibit high contact angles for water, confirming their hydrophobic nature, while allowing significant oil penetration, indicative of oleophilic behavior. The available free volume within the substrates vary from 60\% to 90\% of the total volume of the substrate depending on some parameters of the BD. By combining their water and oil wetting properties with a high availability of volume, the fractal substrates analyzed in this work achieve an efficiency in separating oil from water of nearly 98\%,  which is significantly higher compared to a micro-pillared surfaces made from the same material but lacking a fractal design.
\end{abstract}

\maketitle

\section{Introduction}

 Oil has become the a common pollutant in the world mostly due to oil spill accidents and industry oily wastewater\cite{chan09, pa15} which has lead to an increase concern in studying effective water-oil separation techniques \cite{xue14}.

 The ideal approach for the separation depends on the diameter of the oil droplets contained in water. For stratified oil ($d>20\mu$m)  gravity separation followed by skimming is considered an efficient, low cost,  primary step in water treatment \cite{cher98, zee83}. For emulsion ($d<20\mu$m), on the other hand, these approaches are not effective and follow up steps in treatment, such as chemical emulsification \cite{sun98}, centrifugation \cite{cam06,com90}, heat treatment\cite{strom95} and membrane filtration \cite{kota12, yang2019large}  are often required. These approaches often have high energy and operating costs, limited efficiency and can cause secondary pollution\cite{pa15,duman2024effect}, complicating the cleanup efforts.

In recent years, several materials with specific water and oil wetabilities have been developed for the separation, specially fibrous materials due their connectivity, mechanical flexibility, large specific surface area, and ease of shape manipulation \cite{guo2023bionic}. These materials fall into two categories: oil-removing materials \cite{feng04, gui10, cortese14, zhan20}, which exhibit superhydrophobic/superoleophilic behavior, and water-removing materials \cite{kota12, yang12}, which display superhydrophilic/superoleophobic behavior. Developing water-removing materials is more challenging because most oleophobic materials are also hydrophobic \cite{tuteja07, ahuja08}. As a result, oil-removing materials are more commonly used.

Despite the considerable progress in fabricating such materials, there is still a lack of studies focused on the underlying mechanisms of water-oil separation \cite{chen19}. Thus, more fundamental research is needed to drive further advancements in this field.

In a recent study \cite{gava2021}, we explored two distinct types of substrates to assess how they influence the effectiveness of separating water from oil. Regarding water, it was shown that if a surface repels a droplet of pure water, it will behave similarly when a mixture of oil and water is placed on the same material. In other words, the water's behavior is entirely determined by the surface's hydrophobicity. For oil, however, two factors hinder absorption: the formation of an oil film around the water to minimize the water-gas interface, and the limited available volume for oil absorption, a characteristic of the material from which the substrate is made.

With that in mind, fractal materials present a promising alternative for optimizing water-oil separation. They have been shown to be superhydrophobic \cite{Onda1996, synytska2009wetting, gao2016tunable, iara2023} and offer increased available volume compared to non-fractal materials.

In this study, we explore two types of disordered fractal materials created through ballistic deposition and discuss their application in water-oil separation. Ballistic deposition involves the random arrival of particles onto a substrate, resulting in a structure that typically features interconnected voids, akin to a porous medium, which provides a high volume availability for oil absorption.
Additionally, this generation method yields a very rough substrate, which affects its wetting properties. To characterize these materials, we first examine their wetting characteristics using an energy continuous model, which predicts distinct wetting behavior for both liquids: the substrates exhibit hydrophobic and oleophilic properties. This prediction is validated by Monte Carlo simulations. We also show that these substrates possess a significant amount of available volume for oil absorption due to their porosity. Subsequently, we employ Monte Carlo simulations to evaluate the separation efficiency and the underlying mechanisms influencing this process.

This work is organized as follows: in section  \ref{sec_disordered} it is described the disordered fractal surfaces that will be used throughout the work and the procedure to generate them, in \ref{sec_model} the theoretical continuous model and in \ref{sec_potts} the computational details of the Monte Carlo cellular Potts model. Section \ref{sec_results} presents and discuss the results in terms of efficiency to separate oil and water in those surfaces. In supplementary material (SM), section I we briefly discuss the wetting properties of one example of ordered fractals. Finally, in section \ref{sec_conclusion} we present our conclusions.

\section{Methodology}
\label{sec_method}

Because fractal materials maximize interfacial areas while maintaining a finite volume, they offer a promising alternative for water-oil separation. This property often results in high hydrophobicity and an increased capacity for storing substances, such as oil, compared to non-fractal materials.

Fractal materials can be ordered or disordered. Ordered fractal materials have the advantage of been more easily reproduced and easier to treat analytically. A particular case of ordered material, consisting in a hierarchical pillared surface, was recently studied by Ramos \textit{et. al.}\cite{iara2023}, and it was shown that they present an enhancement of the contact angle when compared to the non fractal version. Despite of that, two issues arise when considering this substrate for water-oil separation: the gain in volume is minimal when the hierarquical level is increased and, depending on the geometric parameters, it may destroy the hydrophobicity of the surface (see Section SM for details). From here, we will focus on the case of disordered fractals, as explained in the following section.

\subsection{Disordered fractal materials: Ballistic Deposition surfaces}
\label{sec_disordered}

A simple model to build disordered fractal materials is the Ballistic Deposition (BD), originally introduced to study colloidal aggregates in the context of available percentage volume~\cite{vold1959numerical,vold1959sediment}.

\begin{figure}
    \centering
    \includegraphics[width=\columnwidth]{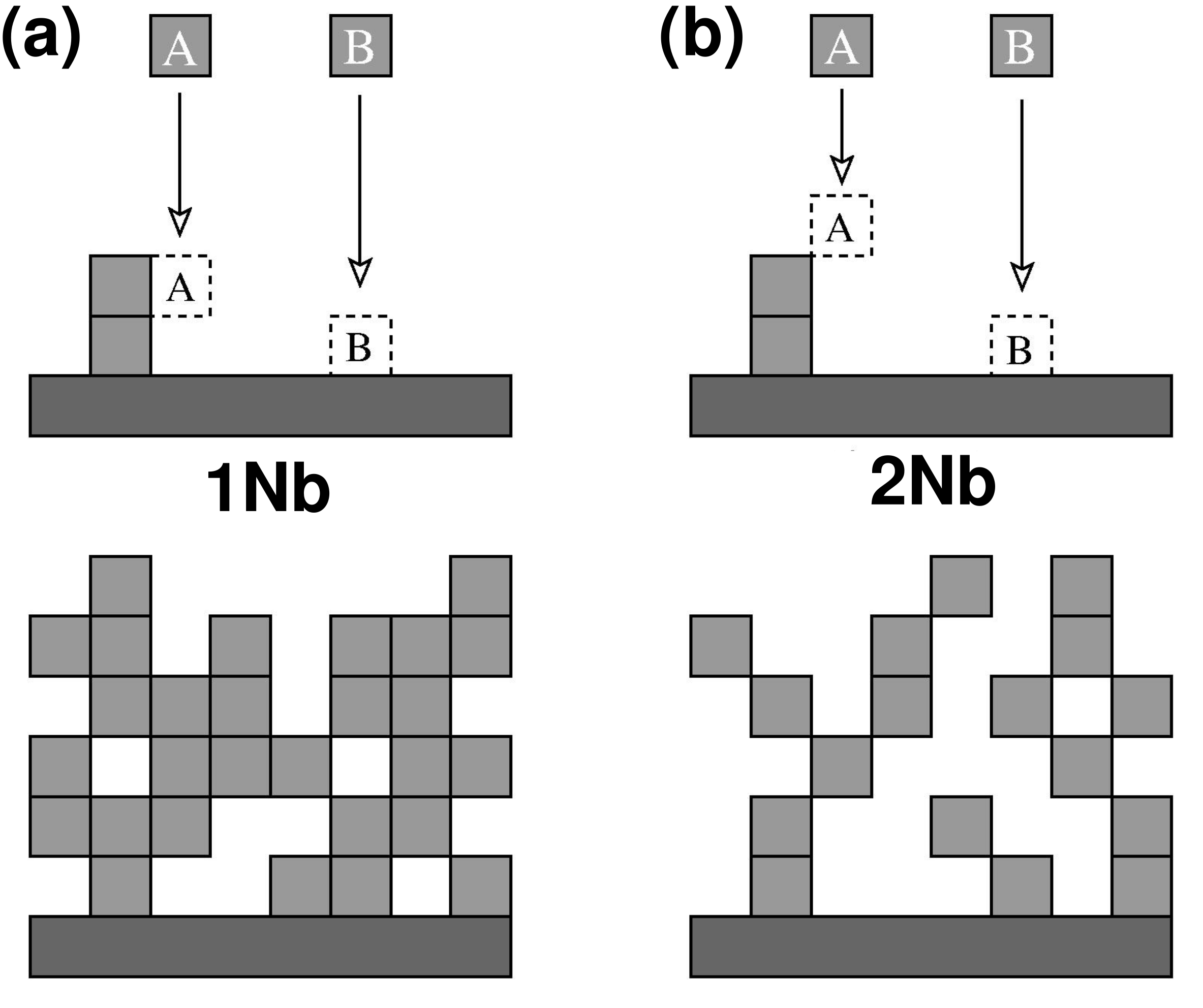}
    \caption{Scheme of first and second neighbors ballistic deposition. a) illustrates the first neighbor sticking rule. Above is shown schematically the sticking rule and below is shown the result after some deposition time. b) illustrates the second neighbor sticking rule. Above is shown schematically how this deposition occurs and below is shown the result after some deposition time.}
    \label{fig_frac_BD}
\end{figure}

In this model, a particle is release from a random position in the x-y plane above the surface. The particle falls in a straight line until it reaches the surface where it sticks, and different sticking rules leads to different materials. Here we use the following two rules: the particles sticks to the first site along its trajectory that have an occupied first neighbor site (refereed as 1Nb) or second neighbor site (refereed as 2Nb). The resulting materials consist of a series of irregularly interconnected vacant spaces that can be regarded as a single irregular continuous porous structure.

Figure \ref{fig_frac_BD} shows two examples of sticking rules where the substrate has only one horizontal dimension $L$ and the vertical axis where the particles are deposited. However, in this work we study the three-dimensional surfaces with a base of size  $L \times L$ and the vertical dimension for particle deposition.   
 The mean height of the surface  $\bar h$  increases linearly as we increase the deposition time $h_0$\cite{fractalgrowthlivro1995}, defined as $h_0 = N/L^2$, where $N$ is the total number of blocks deposed. % and $L$ is the size of the surface in which the material is being deposed. 
 For the materials considered in this work, the porosity of the substrate is defined as the available free volume per unit volume of material \cite{bear2013dynamics}, as illustrated in Figure~\ref{vol_BD}. Clearly, different sticking rules change surface's porosity (and $\bar h$), leading to different amount of vacant volume inside the surface.

\subsection{Continuous model}
\label{sec_model}

The wetting properties of a surface depend not only on its geometry and porosity, but also on its chemistry and the type of liquid interacting with it. To characterize the wetting behavior of the surface, we use a theoretical continuous model to determine the energies involved in creating interfaces when a droplet is placed on top of it. This model treats the droplet as comprising two components: a spherical cap above the surface and a cylindrical volume that penetrates the surface to a defined depth. 

\begin{figure}
	\includegraphics[width=\columnwidth]{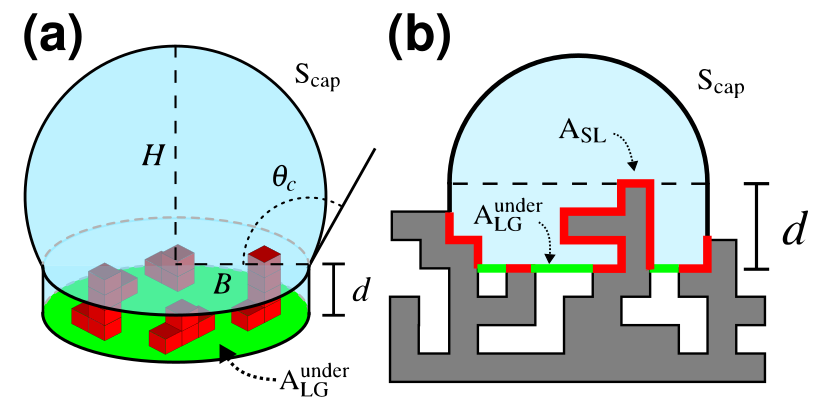}
	\caption{Schema of a droplet deposited in a surface: a) defines the height of the droplet $H$, its base radius $B$, the contact angle $\theta_c$, the penetration depth $d$, the area of the cap $S_{cap}$ and the liquid-gas contact area underneath the penetrated cylinder $A_{LG}^{under}$; b) highlights the liquid-gas and solid-gas contact area, resp. $A_{LG}^{under}$ and $A_{SL}$, in a cross-section of the disordered fractal surfaces.}
	\label{fig_schema}
\end{figure}

It is assumed a  droplet with fixed volume $v_{0} = \frac{4}{3}\pi R^{3}_{0}$, with base radius $B$, cap height $H$ (identified in Figure \ref{fig_schema}), and radius $R$. The droplet can lie in several wetting states, depending on the contact angle $\theta_c$ and the penetrating depth $d$, where $d$ defines how much volume leaves the spherical cap and homogeneously wets the surface. The penetrating depth will depend on a surface height's percentile relation $d = d(Perc)$, detailed in SM. 

The global energy associated with these states is given by  $E_{tot}(d) = \Delta E(d) + E_{g}$, where $E_{g}$ is the gravitational energy and is small in comparison to $\Delta E(d)$ \cite{fernandes2015}.  The $\Delta E(d)$ represents the difference in interfacial energies between pairs of interaction (solid-liquid, solid-gas and liquid-gas) phases before and after placing the droplet on the surface and is given by:

\begin{eqnarray}
    \Delta E(d) &=& A_{SL}(d) (\sigma_{SL}-\sigma_{SG}) +  (S_{cap}+A_{LG}^{\rm under} )\sigma_{LG} , 
    \label{eq_energia}
\end{eqnarray}

where $\sigma_{SL}$, $\sigma_{SG}$ and $\sigma_{LG}$ are the corresponding surface tensions for solid-liquid, solid-gas and liquid-gas interactions, $A_{SL}$ is the contact area between solid and liquid phases,  $A_{LG}^{\rm under}$ is the liquid-gas area underneath the cylinder and $S_{cap}$ is the area of the spherical cap. All the contact areas in Eq.~\ref{eq_energia} depend on the topology of the surface and the penetrating depth $d(Perc)$. Figure \ref{fig_schema}(b) shows a representation of these areas for the disordered fractal material considered in this work.

For disordered surfaces, the contact areas between the surface and the deposited liquid ($A_{SL}$) are challenging to calculate analytically. We address this issue numerically by dividing the surface into pixels and summing the areas where the liquid makes contact with the substrate's pixels.
The model considers the energies for a droplet that homogeneously penetrate the surface until some specific depth  $d$, accounting all the contact area inside the cylinder volume, plus the cap area that is approximated by a semi-sphere $S_{cap}(\theta_c) = 2\pi R^2 (1-\cos(\theta_c))$ for some contact angle $\theta_c$ and the energy is calculated by Eq. \ref{eq_energia}. Details about this model are discussed in the SM.

\subsection{4-spins Cellular Potts model}
\label{sec_potts}

Several numerical approaches are used to study hydrophobic surfaces, including molecular dynamics (MD)\cite{Koishi2011,Wu2009}, lattice Boltzmann methods~\cite{Dupuis2005, Sbragaglia2007}, and the surface evolver~\cite{brakke1992surface, peng2017morphological, xu2022surface}. However, these methods are computationally expensive when applied to mesoscopic systems, which are more suitable for comparison with experimental results. In this context, Monte Carlo (MC) simulations using the cellular Potts model allow for the simulation of larger droplet sizes and are commonly employed to investigate wetting phenomena on textured surfaces~\cite{Lopes2013,Oliveira2011,Mortazavi2013, fernandes2015, gava2021}."

Here we consider a simple cubic lattice in which each state represents one of the components: gas, water, oil, or solid. The Hamiltonian  is given by:

\begin{eqnarray}
H &=& \frac{1}{2} \sum_{\langle{\rm i},{\rm j}\rangle} E_{s_{\rm i},s_{\rm j}}(1-\delta_{s_{\rm i},s_{\rm j}}) + \alpha_w \left( \sum_{\rm i} \delta_{s_{\rm i},1}-V^w_T \right) ^2  \nonumber \\
&+& \alpha_o \left( \sum_{\rm i} \delta_{s_{\rm i},2}-V^o_T \right) ^2 + g \sum_{\rm i} (m_{\rm i} h_{\rm i} \delta_{s_i,1}  + m_{\rm i} h_{\rm i} \delta_{s_{\rm i},2} )
\label{hamil}
\end{eqnarray}

\noindent where the spin $s_i \in \{0,1,2,3\}$ represent gas, water, oil and solid states, respectively.
The first term in Eq. \ref{hamil} represents the energy related to the presence of interfaces between sites of different types. The summation ranges over pairs of neighbors which comprise the 3D Moore neighborhood in the simple cubic lattice (26 sites, excluding the central one), $E_{s_i,s_j}$ is the interaction energies of sites $s_i$ and $s_j$ of different states at interfaces and  $\delta_{s_i,s_j}$ is the Kronecker delta. 

In the second and third terms in Eq. \ref{hamil}, $V^w_T$ and $V^o_T$  are the target water and oil volumes, respectively, the summations are the water and oil volume and the parameters $\alpha_w$ and $\alpha_o$ mimics the liquids compressibility. Thus, these terms maintain the liquids volumes and the desired composition of the droplet constant during the simulation. The last term is the gravitational energy, where $g = 10m/s^2$ is the acceleration of gravity and $m_i$ is the mass of the site. In both the volumetric and gravitational terms, only sites with liquid, $s_i = 1$ or $s_i = 2$, contribute.

\begin{figure}[ht]
    \includegraphics[width=.65\columnwidth]{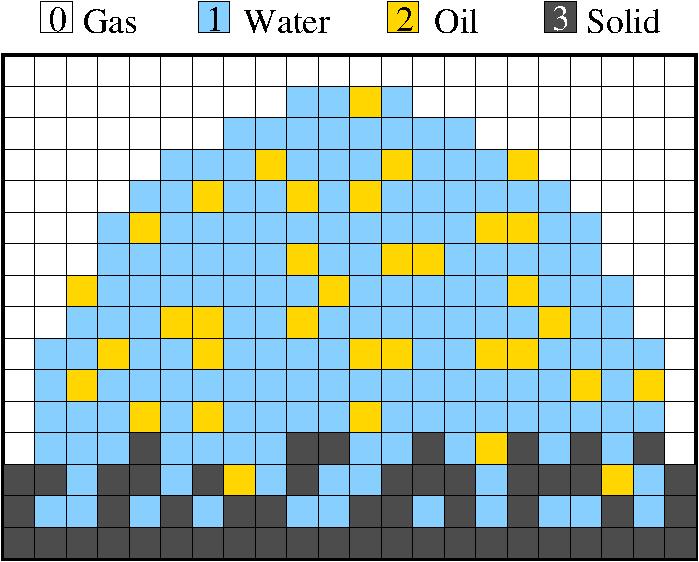}
    \caption{Example of an initial configuration used in simulations of the Monte Carlo Potts Model. }
    \label{ex_CI_simu}
\end{figure}

In our simulations the length scale is such that one lattice spacing corresponds to 1 $\mu$m and the surface tensions values (shown in Table \ref{surface_tention}) are divided by 26, which is the number of neighbors that contributes to the first summation of our Hamiltonian. Therefore, the interfacial interaction energies $E_{s_i,s_j} = A\sigma_{s_is_j}$, with $A=1\mu$m$^2$ are given by $E_{0,1} = 2.70 \times 10^{-9}\mu$J, $E_{0,2} = 1.04 \times 10^{-9}\mu$J, $E_{0,3} = 0.96 \times 10^{-9}\mu$J, $E_{1,2} = 2.06\times 10^{-9}\mu$J, $E_{1,3} = 1.93\times 10^{-9}\mu$J and $E_{2,3} = 0.33\times 10^{-9}\mu$J. The mass existent in a unit cube is $m^w = 10^{-15}$kg for water and  $m^o = 0.77 \times 10^{-15}$kg for oil. We fix $\alpha_w = \alpha_o = 0.01 \times 10^{-9}\mu$J/($\mu$m$)^6$ (these values were chosen to minimize the volume fluctuation while keeping an a suitable acceptation ratio~\cite{gava2021}).

\begin{table}
    \centering
    \begin{tabular}{|c||c|c|c|}
    \hline
         &   Water   & Solid    & Gas \\
    \hline
    \hline
    Oil  & $\sWO=$ 53.5 &  $\sSO=$ 8.6 & $\sGO=$27  \\
  Water  &  ---        & $\sSW=$50.2 & $\sGW=$70    \\
   Solid & ---   & ---       & $\sSG=$ 25  \\
    \hline
    \end{tabular}
    \caption{Surface tensions used in this work in units of mN/m. These values were obtained for $T=25^o$C.}
    \label{surface_tention}
\end{table}

The total run of a simulation is $5 \times 10^{5}$ Monte Carlo steps (MCS), from which the last $2.5 \times 10^{5}$ MCS are used to measure observables of interest. Each MCS is composed by $V^T = V^w + V^o$ number of trial spin flips, where $V^T$ is the volume of the liquid droplet which is composed by a volume of oil $V^o$ and of water $V^w$. A spin flip is accepted with probability $ \text{min}\{1,\text{exp}(- \beta \Delta H)\} $, where $\beta = 1/T$. In the cellular Potts model $T$ acts as noise to allow the phase space to be explored. In our simulations, a value of $T=9$ was used, which allows an acceptance rate of approximately $15\%$ while keeping both water and oil in a liquid state.

The initial wetting state is created using a hemisphere with initial volume $V^T  \approx V_0=4/3\,\pi R_0^3$, due to the discreteness of the lattice. The droplet has $R_0=50 \mu$m  in a cubic box with $L=240\mu m$ or $L=300\mu m$ depending on the surface height. A mixture of oil and water was used. The composition of the droplet is defined by the oil fraction $f_o$, thus, $V^o = f_o V^T$ is the oil volume, by reciprocity, one can also define the water volume  $V^w = f_w V^T$, where $f_w=1-f_o$ is the water fraction. Oil and water sites are randomly distributed in the droplet. In this work 10 distinct surfaces were used for each case studied and all the observable presented here are averages over this surfaces. Figure \ref{ex_CI_simu} shows an example of the initial configurations used.

\section{Results and Discussion}
\label{sec_results}

In this section we present the geometric and wetting properties of the fractal substrates and then explore its efficiency in separating oil and water. 

\begin{figure}
    \centering
    \includegraphics[width=.8\columnwidth]{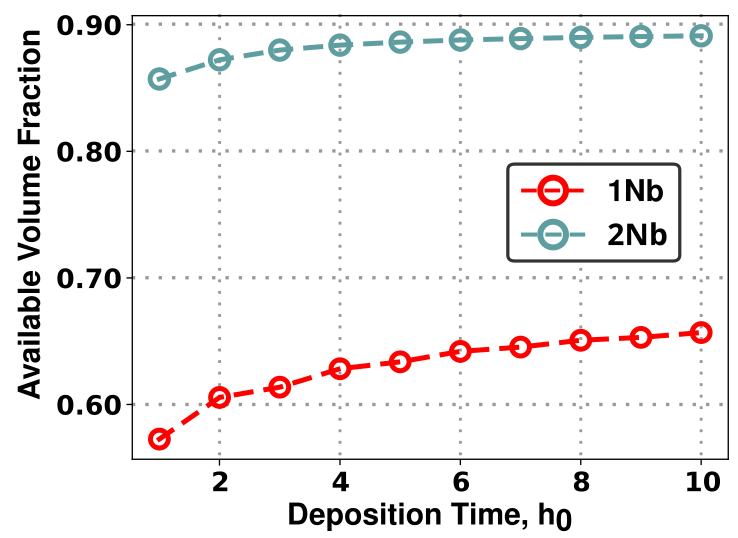}
    \caption{\textbf{Fraction of available volume for Ballistic Depositions} for 1st neighbors (1Nb) and 2nd neighbors (2Nb) sticking rule.}
    \label{vol_BD}
\end{figure}

{\it Available volume for oil absorption.}~
An important factor in oil-water separation is the available free volume within the surface, as it provides capacity for storing the oil mixed with the water~\cite{gava2021}. The available volume, defined as the free volume per unit volume of material, was measured as a function of deposition time $h_0$ for both sticking rules, as shown in Figure \ref{vol_BD}. The results indicate that in both cases, the available volume increases with $h_0$ until it reaches a plateau. Surfaces created using the second neighbors rule (2Nb) achieve nearly 90\% available volume, while surfaces obtained by the first neighbors rule (1Nb) reach 60\%.

%A previous work \sout{shows} \davi{showed} that the behavior of the water in a mix droplet on top of a given surface can be predicted by the behavior of the pure water droplet\cite{gava2021}, therefore, in order to evaluate the hydrophobicity of this materials we used the continuous model for disordered fractals described in \ref{sec_model} and MC simulations of a pure water droplet. \cris{Acho que vou fazer um comentário sobre isso na intro e tirar daqui. Faz mais sentido com o texto como um todo} 

\begin{figure}
    \centering
    \includegraphics[width=\columnwidth]{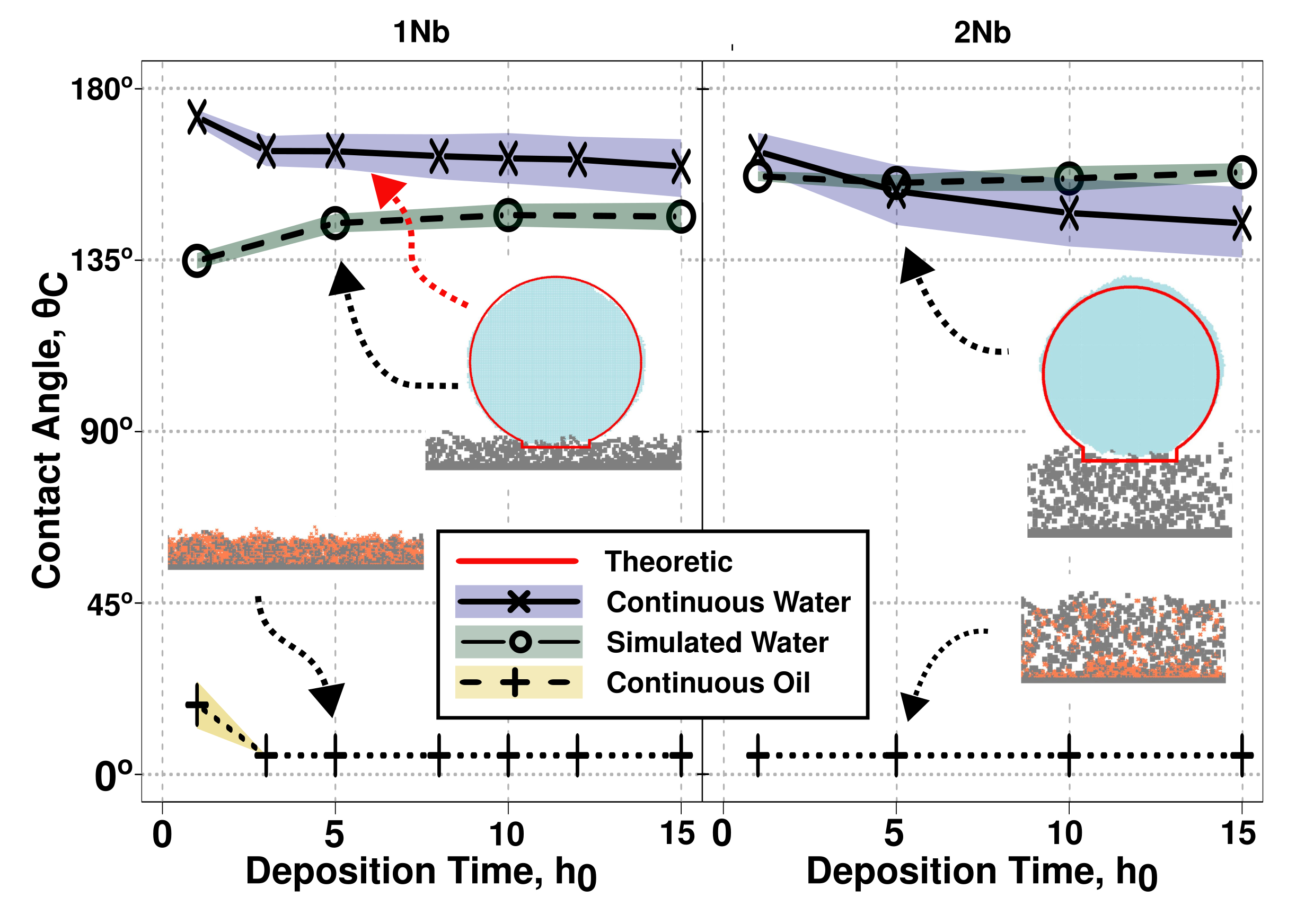}
    \caption{\textbf{Contact angles for water and oil deposed on surfaces of 1st and 2nd neighbors:} Simulated and continuous contact angles as a function of deposition time $h_0$ for the pure water droplet for 1NB (left) and 2Nb (right) sticking rules. The insets show a cross section of the simulated final wetting state for oil (in orange) and water (in blue), and the theoretical continuous final state (in a red line) for water. }
    \label{fig_water_BD}
\end{figure}

{\it Wetting properties of the substrates.~} 
Figure~\ref{fig_water_BD} summarizes the wetting properties of the substrates for different deposition times $h_0$ based on the continuous model and Potts simulations. When a droplet of pure water is placed on the substrates, both methods yield very high contact angles $\theta_C$ —shown as crosses with purple shading for the continuous model and circles with green shading for the Potts model— indicating that the substrates are hydrophobic.

In contrast, pure oil deposition in the continuous model occurs at the minimum allowed angle (approximately $5^{\circ}$), 
with oil penetrating the substrates. This oleophilic behavior is corroborated by Potts simulations, which, although unable to define an angle due to extreme wetting, demonstrate nearly 100\% oil penetration within the surface. 

Overall, both methods consistently indicate that the two types of substrates are hydrophobic and oleophilic, which, combined with the high available volume, is crucial for effective oil-water separation. Their efficiency in promoting this separation are now tested.

{\it Efficiency of substrates in separating oil from water.~} 
To test their efficiency we performed MC simulations for the mix droplet varying the oil concentration for different deposition times. For this purpose we evaluate the composition of the repelled and absorbed phases and the separation efficiency. The first two quantities were analyzed by measuring the percentage of oil and water that remains above, $\upsilon^l_{ab}$, and bellow $\upsilon^l_{b}$, the surface

\begin{eqnarray}
\upsilon^l_{ab} &=& V^l_{ab}/V^l
\label{eq_v_ab}
\\
\upsilon^l_{b} &=& V^l_{b}/V^l
\label{eq_v_b}
\end{eqnarray}

\noindent where the superscription $l$ refers to water, $w$, or oil, $o$, and $V^l$ is the total volume of each liquid in the system, such that $\upsilon^l_{b}+\upsilon^l_{ab} = 1$. The efficiency is defined by

\begin{equation}
\epsilon = \frac{\upsilon^o_{b} + ( 1-\upsilon^w_{b})}{2}.
\label{eq_efficiency}
\end{equation}

The above definition  captures both key elements in oil-water separation: the ability to absorb oil and repel water. These quantities (Eqs. \ref{eq_v_ab}, \ref{eq_v_b} and \ref{eq_efficiency}) were obtained as functions of oil fraction $f_O$ for two distinct deposition times, $h_0=5$ and $h_0=10$, for both sticking rules. 

\begin{figure}
    \centering
    \includegraphics[width=\columnwidth]{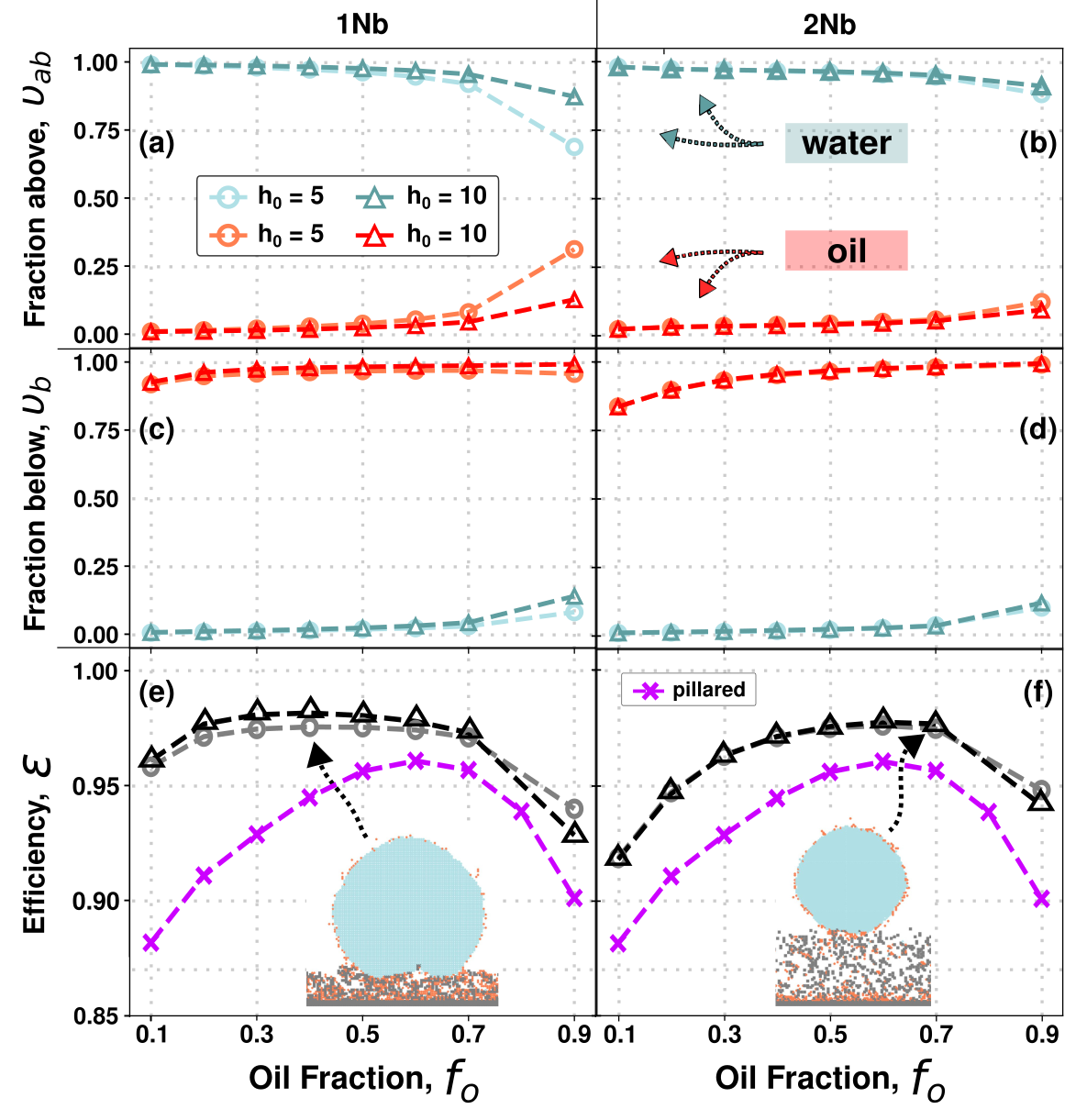}
    \caption{\textbf{Fractions of volume above and below the surface, and oil-water separation efficiency:} percentage of water and oil above the surface for (a) Nb1 and (b) Nb2, percentage of water and oil absorbed by the surface for (c) Nb1 and (d) Nb2  and efficiency for (e) Nb1 and (f) Nb2 as functions of the oil fraction. Insets show the final oil-water separation state obtained by the 4-spin Potts simulation. }
    \label{mix_BD}
\end{figure}

Fig. \ref{mix_BD} summarizes our findings on oil-water separation. In all cases the majority of water remains above  the surface ($\upsilon^w_{ab}\to 1$ and  $\upsilon^w_{b}\to 0$) while the majority of oil is absorbed ($\upsilon^o_{ab}\to 0$ and  $\upsilon^o_{b}\to 1$). However, as the oil fraction $f_o$ is increased, we observe an increase of oil remaining above the surface due to saturation of the surface (Fig. \ref{mix_BD}-(a,b)). Since 1Nb surfaces have lower available volume, this effect is more pronounced for these surfaces and lower deposition times.

The percentage of oil that penetrates the surface also increases with the oil fraction (Fig \ref{mix_BD}-(c,d)). A previous study showed that a film of oil is formed around the water droplet in order to minimize the water-gas interface that is energetic more expensive \cite{gava2021}. The lower the oil fraction, more relevant this oil film is to the percentage of absorbed oil.  Interestingly, this effect is more pronounced for 2Nb sticking rule and  can be explained by the oleophilicity of these surfaces. The energy associated to a column of oil penetrating the surface is $\Delta E \approx (\sSO-\sSG)f_s$, where $f_s$ is the solid fraction of the surface. Since the solid fraction is bigger for 1Nb surfaces and $\sSG > \sSO$, the energy involved in this process  is smaller for 1Nb then for 2Nb surfaces, what make 1Nb surfaces more oleophilic than the 2Nb ones.  

These behaviors result in efficiencies $\epsilon$ shown in Fig. \ref{mix_BD}-(e,f). For both sticking rules the efficiency increases as $f_o$ grows, until it reaches a maximum  of approximately 98$\%$ and decreases again for higher oil fraction values. This decrease can be explained partially due to the saturation of the surface and partially by  the MC  that allows for some water sites to appear under the surface. This later effect being more relevant as the amount of water in the droplet is decreased. 

For 1Nb surfaces the maximum value of efficiency is reached for very low values of $f_o$ and it remains approximately as a plateau until it decreases for $f_o=0.90$. For 2Nb surfaces the maximum efficiency occurs only for $f_o = 0.50 - 0.60$ due to the lower oleophilicity discussed above.

Therefore, the gain in available volume obtained trough the 2Nb sticking rule is achieved by losing some solid contact areas resulting in a less oleophilic surface. This loss facilitates other phenomenon, less advantageous for water-oil separation, such as the oil film around the water droplet above the surface.

Fig. \ref{mix_BD}-(e,f) also compares the efficiency $\epsilon$ of a pillared substrate (shown in purple), as studied in \cite{gava2021}. While the pillared substrate shares the same chemical properties as the fractal surfaces analyzed in this work and has been shown to be hydrophobic and oleophilic, its non-fractal nature results in lower efficiency for oil-water separation across all values of  $f_o$.

Some experimental works also show very high efficiency in the oil-water separation, as for example in Refs\cite{gu2014robust,singh2016fabrication,wang2017novel,su2019rubber}. The comparison with the experimental works is however tricky because of the difference in the definitions of the efficiencies, which usually only takes into account the purity of the retained water or the oil that passes. Moreover, most of the experimental studies are made with oil droplets that are big enough to react the gravity.
 
\section{Conclusions}
\label{sec_conclusion}

%\cb{faltando i) comparacao com experimentos (mesmo que usando outras definicoes de eficiencia), ii) critica quanto ao fato de que estamos usando uma esponja. iia) Podemos sugerir que a superficie seja apenas uma membrana que deixe o oleo passar fazendo o processo mais de uma vez (tipo wang 2017?) ou iib) pensar em retirar o oleo da superficie  }

This work shows that fractal surfaces with a high level of porosity, more specifically fractal surfaces grown through Ballistic Deposition (BD) process , are good candidates to serve as oil-water separation surfaces.
We employed a continuous theoretical model designed for disordered surfaces and a 4-spin Potts model simulation to show that these BD-produced surfaces exhibit distinct wetting properties for each liquid, being hydrophobic and oleophilic. Additionally, their substantial free space addresses the oil saturation issue discussed in \cite{gava2021}.

We evaluated the oil-water separation efficiency across different deposition times ($h_0$) and oil fractions. Our simulations indicate that efficiency ($\epsilon$) can reach 98\% under certain parameter ranges, significantly surpassing that of non-fractal substrates made from similar materials.

We investigated two types of fractal substrates generated using different stick rules in the BD process, resulting in varying  available volumes to absorb oil. Notably, the substrate with greater volume exhibited slightly lower efficiency in separating oil from water. This was attributed to its reduced oleophilicity, as it has fewer contact areas between the solid and liquid.

A key limitation of this work is that the substrates behave like sponges for oil, requiring removal for potential reuse. However, we hypothesize that these surfaces could function as membranes for separating oil from water, particularly in situations where the oil quantity is substantial enough to be influenced by gravity. The possibility and efficiency of the utilization of a substrate as a membrane and a sponge was previously studied in Ref. \cite{zhou2023bio}. Similarly, passing the oil-water mixture through the membrane studied here multiple times could enhance separation efficiency .

\section{Supplementary material}

See supplementary material for a brief discussion on ordered fractals and a detailed description of the continuous model for disordered fractals.

\begin{acknowledgments}
We thank the Brazilian agency CAPES and CNPq for the financial support. 
\end{acknowledgments}

\section*{Data Availability Statement}
The data that supports the findings of this study are available within the article and its supplementary material. 

\section*{Author Contributions}
C.G. and C.B. designed research. C.G., D.L., I.P.S.R. and C.B performed research, contributed new reagents/analytic tools, analyzed data, and wrote the paper.

\section*{Conflict of Interest Statement}
The authors have no conflicts to disclose.

%\nocite{*}
\bibliography{biblioteca}% Produces the bibliography via BibTeX.

\end{document}